\documentclass{aa}
\usepackage{times}
\usepackage{natbib}
\usepackage[dvips]{graphicx}
\bibpunct{(}{)}{;}{a}{}{,}

\def\4U{4U 1543-475}
\def\XMM{{\em XMM--Newton}}
\def\RXTE{{\em RXTE}}

\begin{document}

\title
{\XMM~observation of \4U: the X--ray spectrum of a stellar--mass black--hole at low luminosity}

\author{N. La Palombara \& S. Mereghetti}

\institute {{Istituto di Astrofisica Spaziale e Fisica Cosmica -
Sezione di Milano ``G.Occhialini'', CNR, via Bassini 15, I-20133
Milano, Italy}}

\offprints{N. La Palombara, nicola@mi.iasf.cnr.it}

\date{Received / Accepted}

\authorrunning{N. La Palombara \& S. Mereghetti}

\titlerunning{\XMM~observation of \4U}

\abstract{We report the results of an observation of the galactic black--hole
binary \4U performed by \XMM~on 2002 August 18, about two months after the start
of an outburst detected by \RXTE.  Despite the relatively low flux of the
source, corresponding to a luminosity $L_{X}\sim 4\times 10^{34}$ erg
s$^{-1}\sim 10^{-5} L_{EDD}$, we could obtain a good quality spectrum thanks to
the high throughput of the \XMM~{\em EPIC} instrument.  The spectrum is well fit
by a power law with photon index $\Gamma$=1.9--2 without any evidence for iron
emission lines or for thermal emission from an accretion disk.  We could
estimate an upper--limit on the disk bolometric luminosity as a function of the
colour temperature:  it is always lower than $\sim10^{33}$ erg s$^{-1}$, i.e.
less than 10 \% of the source total luminosity.  Finally, we evaluated that the
disk colour temperature must satisy the condition $kT_{col}<$ 0.25 keV in order
to obtain an acceptable value for the disk inner radius.
}

\maketitle

\section{Introduction}

\4U is a recurrent X-ray transient containing a black hole.  It was discovered
during an outburst in 1971 \citep{Matilsky1972} and observed in outburst again
in 1983 \citep{Kitamoto1984}, in 1992 \citep{Harmon1992} and in 2002
\citep{Park2004, Kalemci2004}.  In the decade--long quiescence periods the flux
is lower than 0.1 $\mu$Crab, while the source brightens by a factor greater than
$2\times10^{7}$ in outburst \citep{Garcia2001}.  In the latest outburst, between
June and July 2002, the measured flux reached a peak value of 4.2 Crab in the
2--12 keV energy band, comparable to the peak intensities observed in the three
previous outbursts \citep{TanakaLewin1995}.  The source light curve along the
whole outburst showed the characteristic shape of the `classic' X--ray novae,
i.e.  a fast rise to the outburst peak followed by an exponential decay, with
{\em e}--folding decay time of $\simeq$ 14 days \citep{McClintockRemillard2003}.
The spectrum was soft and dominated by the emission from the accretion disk:
its continuum part was fit with a multi--colour disk blackbody (kT$_{max}$ =
1.04 keV) and a power--law (photon index $\Gamma \sim 2.7$).

The optical counterpart of \4U, IL Lupi, has been classified as spectral type
A2V \citep{ChevalierIlovaisky1992}.  During the X--ray outbursts it brightens by
$\simeq$ 1.8 mag \citep{vanParadijsMcClintock1995}, but during the quiescent
periods it was possible to perform detailed studies.  This allowed to derive the
orbital period $P_{orb}$ = 26.8 hr, the black--hole mass $M_{1}=9.4\pm2.0
M_{\odot}$, the secondary star mass $M_{2}=2.7\pm1.0 M_{\odot}$, the source
distance $d=7.5\pm1.0$ kpc and the orbit inclination $i = 20.7\pm1.0^{\circ}$
\citep{Orosz1998,Orosz2003}.

Here we present the results of an \XMM~observation obtained after the most
recent outburst of \4U.  The spectrum obtained with the {\em EPIC} instrument
has an unprecedented statistics for a galactic black--hole at this level of
luminosity, allowing for the first time to investigate the spectral details of a
black--hole binary approaching the quiescent phase.

\section{Observations, data reduction and spectral analysis}

\begin{table*}[t]
\begin{center}
\caption{Best--fit parameters for a power--law model in the case of the
PN, MOS2 and PN+MOS2 data.}\label{fit}
\footnotesize{
\begin{tabular}{ccccccc} \hline
Instrument	& $N_{H}$		& Photon		& Normalization at 1 keV			& $\chi^{2}_{\nu}$	& d.o.f.	& $f_{X}$ (0.3--10 keV)$^{a}$	\\
	& ($\cdot 10^{21}$ cm$^{-2}$)	& Index		& (ph keV$^{-1}$ cm$^{-2}$ s$^{-1}$)	&			&	& (erg cm$^{-2}$ s$^{-1}$)	\\ \hline
PN		& 3.6 $\pm$ 0.2		& 1.92 $\pm$ 0.04	& (9.9 $\pm$ 0.5) $\times 10^{-4}$		& 1.104			& 297	& (3.7 $\pm$ 0.2) $\times 10^{-12}$	\\
MOS2		& 4.5 $\pm$ 0.3		& 1.99 $\pm$ 0.07	& (11.2 $\pm$ 0.9) $\times 10^{-4}$		& 1.131			& 146	& (3.7 $\pm$ 0.3) $\times 10^{-12}$	\\
PN+MOS2	& 3.8 $\pm$ 0.2		& 1.94 $\pm$ 0.04	& (10.3 $\pm$ 0.3) $\times 10^{-4}$		& 1.069			& 442	& (3.7 $\pm$ 0.1) $\times 10^{-12}$	\\ \hline
\end{tabular}}
\end{center}
\begin{small}
NOTE - Errors are at 90 \% c.l. for a single interesting parameter; $^{a}${Absorbed flux}\\
\end{small}
\vspace{-0.5 cm}
\end{table*}

\4U was observed by \XMM~on 2002 August 18, between 12:30 and 20:42 UT.  The
source was on--axis and the observation lasted about 29.5 ks.  The three {\em
EPIC} cameras \citep{Turner2001, Struder2001} were all active during the
observation:  the PN, MOS1 and MOS2 instruments were operated in {\em Small
Window}, {\em Timing Uncompressed} and {\em Full Frame} mode, respectively.  For
all of them, the Thin filter 1 was used.

We used the version 5.4.1 of the \XMM~{\em Science Analysis System} ({\em SAS}) to process
the event files.  After the standard processing pipeline of the data, we looked for
possible periods of high background due to soft proton flares with energies
less than a few hundred keV.  For the two MOS cameras this was done by inspecting the
light curves of the events with energy above 10 keV detected in the peripheral CCDs and
with pixel patterns from 0 to 4.  In this way, we selected only those background events
whose high energy is distributed in one or two pixels at most, which presumably are due to
low energy protons focused on the focal plane by the telescope mirrors.  These light
curves showed a large increase of the count--rate (up to a factor of $\sim$ 40) during the
last 10 ks of the observation.  We decided to use for our analysis of \4U only time
intervals with count--rates below 0.7 c s$^{-1}$ for this type of events, yielding an effective
observing time of 19.9 ks.

Since the PN camera was operated in {\em Small Window} mode, no data were taken with the
peripheral CCDs.  Therefore, the data cleaning was based on the events in the central CCD,
with energies above 10 keV and pattern 0\footnote{the PN pixels are larger than the MOS
ones; therefore, no multi--pixel events are expected by the soft--protons}.  Also in this
case we observed a large increase of the count--rate in the last part of the observation:
we selected time periods with count--rates lower than 0.07 c s$^{-1}$, thus obtaining
an effective exposure time of 13.2 ks\footnote{this is smaller than the MOS one due to the
reduced live--time of the Small Window mode used for the PN}.

In order to study the temporal behavior of the source, we accumulated its PN
light curve in the whole energy range 0.2--10 keV and in the two sub--ranges
0.2--2 and 2--10 keV, using an extraction radius of 30$''$.  The corresponding
background curves were extracted over the part of the CCD area not affected by
the wings of the source {\em Point Spread Function}.  The background--subtracted
light curve in the whole energy range shows some variability.  The light curve
binned at 1000 seconds gives a reduced $\chi^{2}_{\nu}$=2.6 when fitted to a
constant value and shows variations up to $\sim$30 \% around the average level.
The hardness ratio between the two sub--ranges is consistent with a constant
value, indicating that no significant spectral variations took place along the
observation.  Therefore we performed a single spectral analysis on the whole set
of data.

\begin{figure}[h]
\centering
\resizebox{\hsize}{!}{\includegraphics[angle=-90,clip=true]{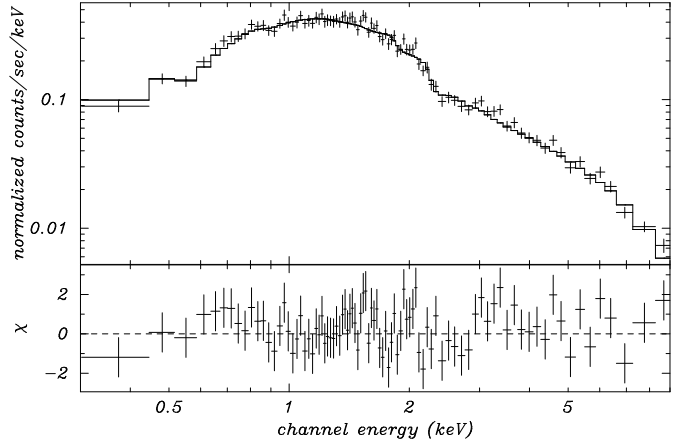}}
\caption{Comparison of the PN spectra of \4U with the best--fit power--law model.  In the
lower panel the residuals (in units of $\sigma$) between data and model are shown.}\label{spectrum}
\end{figure}

For the PN spectra of both the source and the background, we considered the same
focal plane areas used for the light--curves.  We used the {\em SAS} software to
calculate the applicable response matrix.  The accumulated source spectrum was
rebinned in order to have at least 30 counts in each energy bin.  For the
spectral analysis we used {\em XSPEC} 11.2.0.  The source spectrum was fitted
with an absorbed power--law model, in the energy range 0.3--8.5 keV:  the
best--fit parameters are reported in Tab.~\ref{fit}.  In Fig.\ref{spectrum} we
show the PN spectrum and the best--fit power--law model.

Fits with simple thermal emission models provided unacceptable results.  For
example, with a bremsstrahlung model we obtained $\chi^{2}_{\nu}$ = 1.234, while
a multi--temperature blackbody accretion disk ({\em diskbb} in {\em XSPEC})
gave $\chi^{2}_{\nu}$ = 1.907, always with 297 d.o.f.

For the MOS2 camera\footnote{we have not used the MOS1 data since the spectral
response in Timing Mode is not yet well calibrated} we applied the same source
extraction radius of the PN camera (30$''$), while, thanks to the larger size of
the MOS CCDs, we could accumulate the background spectrum using a circular area
of 2.5$'$ radius.  Also in this case the spectrum was rebinned with a minimum of
30 counts in each energy bin and fitted over the 0.3-8.5 keV energy range.  The
fit with a power--law yielded a comparable photon--index but a higher hydrogen
column density than the PN one (Tab.~\ref{fit}).  This discrepancy is likely due
to the uncertainties still remaining in the calibration of the two instruments,
especially at low energies \citep{Kirsch2004}.  We also fitted the PN and MOS2
data simultaneously, after including a systematic error of 5\% in the spectral
bins.  Again a satisfactory fit was obtained with a power--law model (see
Table~\ref{fit}) while thermal models gave worse $\chi^{2}$ values.

In conclusion, although the exact value of the absorption is still subject to
some uncertainty\footnote{note, however, that this is one of the best
measurements of $N_{H}$ for this source obtained so far; our value is comparable
with the value of ($4.26\pm0.15$)$\times 10^{21}$ cm$^{-2}$ obtained by
\citet{VanDerWoerd1989} with {\em EXOSAT} data}, all our data point to a
power--law spectrum with photon index $\sim$2.  In the following analysis, aimed
to assess the possible presence of additional components in the spectrum, we
will consider only the PN data, due to their higher statistics ($\sim$12000
source counts are detected in the PN compared to $\sim$5600 in the MOS2).  We
checked that similar results would be obtained using the MOS2 data, although
with looser upper limits (usually a factor 2--5 times higher).

To look for a possible Fe K$_{\alpha}$ emission line, we tried
to improve the spectral fit by adding a gaussian component at four
different fixed energies between 6 and 7 keV, with $\sigma$ between 0 and
0.5 keV.  In all the cases we found no evidence for it:  the gaussian
normalization was always consistent with zero.  The upper limit on the
line equivalent width is always below $\sim$ 0.25 keV at a 3 $\sigma$
confidence level (Tab.~\ref{narrowline}).

\begin{table}
\begin{center}
\caption{Upper limits (3 $\sigma$ confidence level) on the Equivalent Width of an
emission line (gaussian model), at various energies and widths.}\label{narrowline}
\footnotesize{
\begin{tabular}{ccccc} \hline
E (keV)		& 6		& 6.4		& 6.7		& 7		\\ \hline
$\sigma$ (keV)	& U.L. (eV)	& U.L. (eV)	& U.L. (eV)	& U.L. (eV)	\\ \hline
0		& 173		& 117		& 37		& 97		\\
0.1		& 209		& 130		& 48		& 99		\\
0.2		& 255		& 150		& 71		& 91		\\
0.3		& 262		& 161		& 90		& 97		\\
0.4		& 254		& 173		& 113		& 111		\\
0.5		& 235		& 176		& 138		& 134		\\ \hline
\end{tabular}}
\end{center}
\vspace{-0.5 cm}
\end{table}

Although not formally required by the data, which are satisfactorily fit by the
power--law model, we investigated the possible presence of an additional
multi--colour disk emission \citep{Mitsuda1984, Makishima1986}.  To this aim we
added a {\em disk blackbody} component and obtained $N_{H} =
(4.4\pm0.7)\times10^{21}$ cm$^{-2}$, $\Gamma=1.94\pm0.05$,
$K_{PL}=(1.0\pm0.1)\times 10^{-3}$, $kT_{col} = 0.16\pm 0.03$ keV and
$K_{DBB}=141\pm289$ as best--fit parameters, with $\chi^{2}$/d.o.f.  =
325.5853/295.  This result would imply an absorbed flux $f_{DBB}^{0.3-10} =
8.44\times 10^{-14}$ erg cm$^{-2}$ s$^{-1}$ for the thermal component (i.e.
$\sim$ 2\% of the total flux), but the large error on its normalization clearly
demonstrates that this component is only marginally present.  The F--test proves
that the fit quality improves only at a 0.3 $\sigma$ level.

\begin{figure}[h]
\centering
\resizebox{\hsize}{!}{\includegraphics[angle=-90,clip=true]{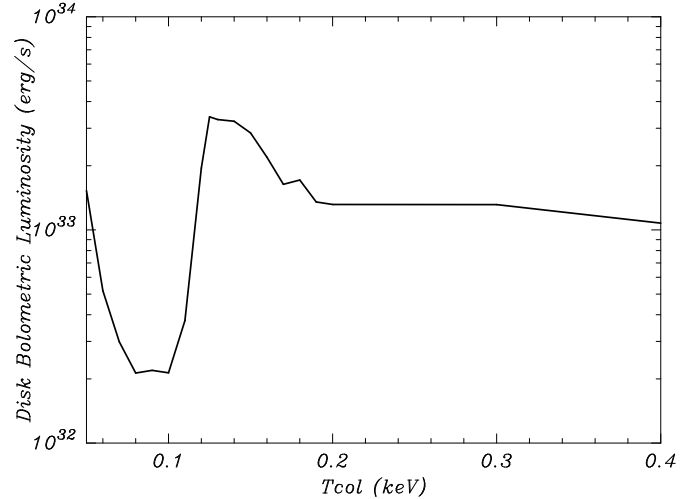}}
\caption{3 $\sigma$ upper--limit on the bolometric luminosity of the multi--colour disk
as a function of the colour temperature.}\label{contour}
\end{figure}

We have therefore computed an upper limit to the disk emission component.  This
is shown in Fig.~\ref{contour}, where the upper--limit on the disk bolometric
luminosity is shown as a function of the colour temperature $T_{col}$ (assuming
a source distance d = 7.5 kpc).  The power--law best--fit parameters imply an
unabsorbed flux $f_{X}=(5.83\pm 0.29) \times 10^{-12}$ erg cm$^{-2}$ s$^{-1}$ in
the energy range 0.3--10 keV, which corresponds to $L_{X}=(3.92\pm 0.20) \times
10^{34}$ erg s$^{-1}$.  This means that any possible disk contribution is well
below 10 \% of the total source luminosity.

\section{Discussion and conclusions}

\begin{figure}[h] \centering
\resizebox{\hsize}{!}{\includegraphics[angle=-90,clip=true]{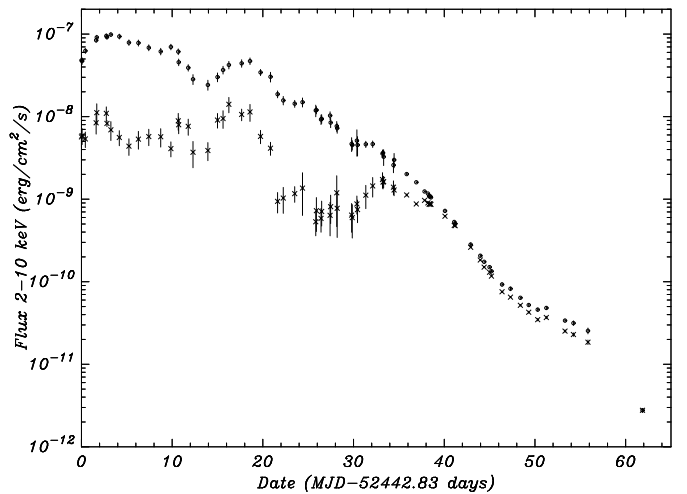}}
\caption{X--ray flux observed by \RXTE~and \XMM~during the 2002 outburst of \4U.
The open circles refer to the total source flux, the crosses to the
power--law component. The last point to the right is our \XMM~total flux}\label{xtelc}
\end{figure}

The \XMM~data reported here were obtained just two months after the start of the source
outburst observed by {\em Rossi--XTE} (\RXTE) on MJD 52445 (2002 June 20).  The
source was monitored by \RXTE~during all this period:  \citet{Park2004} analysed
the data collected between MJD 52442 and 52477 (i.e.  before the transition to
the {\em low--hard} state), while \citet{Kalemci2004} considered the data
between MJD 52464 and 52498 (i.e.  during and after the transition to the {\em
LH} state).  Both \citet{Park2004} and \citet{Kalemci2004} used the sum of a
power--law and a multi--colour disk blackbody to fit the continuum part of the
spectra, finding that the photon index decreased from 3.85 (at the outburst
peak) to 2.21 (just $\sim$ 6 days before the \XMM~observation), while the
temperature at the inner disk edge decreased from 1 to 0.35 keV.  Using the
best--fit parameters reported by these authors, we computed the 2--10 keV fluxes
as a function of time for the total emission and for the power law component
only, which are shown in Fig.\ref{xtelc}.  The last point in the figure
indicates our \XMM~flux value, converted to the same energy range.  Even if this
flux is below the extrapolation of the last \RXTE~measurements, the plot shows
that it is in general agreement with the overall decreasing rate measured by
\RXTE.

The luminosity observed with \XMM~corresponds to $\sim1.5\times 10^{-5}$ of the
{\em Eddington} luminosity, still well above the typical quiescent level of BH
candidates $L_{q}\sim 10^{-8.5} L_{EDD}$ \citep{McClintockRemillard2003}, where
we assumed d = 7.5 kpc and M = 9.4 M$_{\odot}$.  At the time of the \XMM~
observation (i.e.  two months after the outburst), 4U 1543-475 was not yet in
quiescence but still in the low--hard state, which is characterized by 1.5$<
\Gamma <$ 2.1 and a power--law flux contribution higher than 80 \% of the total
flux.

According to \citet{Kalemci2004} some emission from an accretion disk is still
required by the last \RXTE~spectra of \4U obtained six days before the
\XMM~observation reported here.  Their data cannot yield a disk temperature for
such a faint and soft component.  Therefore Kalemci et al.  fix it at the value
of 0.35 keV.  Assuming d=7.5 kpc and i=21$^{\circ}$, for such a temperature our
upper limit on the disk blackbody normalization corresponds to a disk {\em
colour} radius of 0.08 $R_{g}$ (where $R_{g}=GM/c^{2}$ is the BH gravitational
radius).  The value of $r_{col}$ obtained from the best fitting procedure of the
multi--colour disk blackbody model systematically underestimates the actual disk
radius \citep{ST95, MFR2000}.  In fact $r_{in}=\eta \times g\times
f^{2}_{col}\times r_{col}$, where $\eta$ = 0.6--0.7 is the ratio between the
inner radius and the effective radius (i.e.  the radius at which the temperature
of the disk peaks), $g$ = 0.7--0.8 takes into account the general relativistic
corrections and $f_{col}=T_{col}/T_{eff}$ is the spectral hardening factor,
whose value can be about 3 for low accretion rates.  If we consider the maximum
value for all the parameters, from $r_{col}=0.08 R_{g}$ we obtain $r_{in}\sim0.4
R_{g}$.  This value is incompatible with the innermost stable circular orbit
around a black--hole, which is $R_{ISCO}=6 R_{g}$ for a Schwarzschild BH and
$R_{ISCO}=R_{g}$ for a Kerr BH \citep{ShapiroTeukolsky1983}.  Note that the same
would be true for lower normalizations at the same temperature.  Only for disk
temperatures $kT_{col}<$ 0.25 keV our normalization upper limits give
acceptable values for the inner disk radius.  For example, for $kT_{col}$ = 0.2
keV we obtain $r_{col}=0.43 R_{g}$, which corresponds to $r_{in}\sim 2.2 R_{g}$.
A similar conclusion was noted in a short report on this XMM observation by
\citet{Miller2003}, but contrary to these authors we do not find a significant
improvement by the inclusion of the disk component in the fit to the spectrum of
\4U.

In conclusion the observation reported here provides one of the best quality
spectra for a black hole X--ray binary accreting at a low level, but still far
from the quiescent state.  The featureless power law energy spectrum does not
give any evidence for the presence of an accretion disk.  If a disk is present,
it is constrained to have a luminosity of (2-30)$\times 10^{32}$ erg s$^{-1}$
and a temperature smaller than 0.25 keV.  This is an interesting result since
little is known on the possible presence and the properties of accretion disks
around black--holes at these relatively low luminosities.  Even if a
multi--colour disk is not strictly required in the hard state from the
theoretical point of view, there are some arguments in favour of it:  on one
hand, most common accretion models foresee the presence of a disk with a large
(i.e.  $\ge 100 R_{g}$) inner radius \citep{Esin1997, McClintockRemillard2003};
on the other hand, a soft X--ray excess, which can be modeled with a large and
cool accretion disk, has been observed in GX 339-4 \citep{Wilms1999} and XTE
J1118+480 \citep{McClintock2001, Frontera2003}, albeit at higher luminosities
than that reported here for \4U.

\begin{acknowledgements}
This work is based on observations obtained with XMM-Newton, an ESA science
mission with instruments and contributions directly funded by ESA Member States
and NASA.
\end{acknowledgements}

\bibliographystyle{aa}
\bibliography{biblio}

\end{document}